\documentclass[debug,overfull]{epl}
\usepackage{graphicx}
\usepackage{dcolumn}
\usepackage{bm}

\title{Novel low-energy collective excitation at metal surfaces}
\author{V. M. Silkin\inst{1}, A. Garc\'\i a-Lekue\inst{2}, J. M.
Pitarke\inst{1,2}, E. V. Chulkov\inst{1,3}, E. Zaremba\inst{4}, \and P. M.
Echenique\inst{1,3}}
\institute{\inst{1} Donostia International Physics Center (DIPC) and Centro
Mixto CSIC-UPV/EHU - Manuel de Lardizabal Pasealekua, E-20018 Donostia, Basque
Country, Spain\\
\inst{2} Materia Kondentsatuaren Fisika Saila, Zientzi Fakultatea,
Euskal Herriko Unibertsitatea - 644 Posta kutxatila, E-48080 Bilbo, Basque
Country, Spain\\
\inst{3}Materialen Fisika Saila, Kimika Fakultatea, Euskal Herriko
Unibertsitatea - 1072 Posta kutxatila, E-20018 Donostia, Basque Country,
Spain\\
\inst{4}Department of Physics, Queen's University - Kingston, Ontario, Canada
K7L 3N6}


\begin{document}

\maketitle

\begin{abstract}
A novel collective excitation is predicted to exist at metal surfaces where a
two-dimensional surface-state band coexists with the underlying
three-dimensional continuum. This is a low-energy acoustic plasmon with
linear dispersion at small wave vectors.  Since new modern spectroscopies are
especially sensitive to surface dynamics near the Fermi level, the existence
of surface-state induced acoustic plasmons is expected to play a key role in a
large variety of new phenomena and to create situations with potentially new
physics.
\end{abstract}

The long-range nature of the Coulomb interaction between electrons in a
three-dimensional (3D) metal is known to yield collective behaviour,
manifesting itself in the form of plasma oscillations \cite{pines}.
These collective charge-density oscillations are a fundamental property of
metals, and are known to be of basic importance in the interpretation of a
large variety of experiments.

In 1957 Ritchie introduced the idea of surface plasmons \cite{ritchie}, which
are collective oscillations bound to a metal surface. In the long-wavelength
limit, the energy of the surface plasmon is given by
$\hbar\omega_s=\hbar\omega_p/\sqrt{2}$, where $\omega_p=(4\pi n
e^2/m_e)^{1/2}$ is the bulk-plasma frequency, $n$ being the electron density
and $m_e$ the electron mass. The concept of surface
plasmons has played a significant role in a variety of areas of fundamental
and applied research, from surface dynamics \cite{ishida,rocca} to
surface-plasmon microscopy \cite{nature1}, surface-plasmon resonance
technology \cite{nature3}, and a wide range of photonic applications
\cite{nature4,science1,science2}.

Another collective excitation arises in a two-dimensional (2D) gas of electrons
confined to a plane. In this case, the 2D plasmon has energy
$\hbar\omega_{2D}=\hbar(2\pi n^{2D}e^2q/m_e)^{1/2}$ ($n^{2D}$ is the 2D
electron density), which goes to zero as $q^{1/2}$ as the 2D wave number $q$
tends to zero \cite{stern}. The 2D plasmon was first observed in artificially
structured semiconductors \cite{2d1} and more recently in a metallic
surface-state band on a silicon surface \cite{2d2}. 

Since the typical energy $\hbar\omega_p$ of a three-dimensional (3D) plasmon
is a few electronvolts, thermal excitation of bulk and surface plasmons is
improbable and the electronic properties near the Fermi level cannot be
influenced by these excitations. Plasmons in a 2D electron gas have lower
energies, but they still do not affect electron-hole (e-h) and phonon dynamics
near the Fermi level due to their square-root dependence on the wave
vector. Much more effective than ordinary 3D or 2D plasmons in mediating,
e.g., superconductivity would be the so-called acoustic plasmons with
sound-like long-wavelength dispersion, which have spurred over the years a
remarkable interest and research activity \cite{tosi}.

Pines originally suggested that acoustic plasmons could be realized in the
collective motion of a system of two types of electronic carriers
\cite{pines2}. The possibility of having a longitudinal acoustic mode in a
metal-insulator-semiconductor structure was anticipated by
Chaplik \cite{chaplik}. Chaplik considered a simplified model in which a 2D
electron gas is separated from a semi-infinite metal. He found that the
screening of valence electrons in the metal changes the 2D plasmon energy from
the square-root behaviour to a linear dispersion, which was also discussed by
Gumhalter \cite{gum} in his study of transient interactions of surface-state
e-h pairs at metal surfaces. However, acoustic plasmons were only expected to
exist for spatially separated plasmas, as pointed out by Das Sarma and
Madhukar \cite{sarma}. Acoustic plasma oscillations were observed in
two-dimensionally confined and spatially separated multicomponent structures
such as quantum wells and heterojunctions \cite{olego}, and were
then proposed as possible candidates to mediate the attractive interaction
leading to the formation of Cooper pairs in high-$T_c$ superconductors
\cite{ruvalds,kresin}.

In this Letter, we show that in metals with a partially occupied surface-state
band the dynamical screening at the surface provides a mechanism
for the existence of a {\it new} acoustic collective mode whose energy
exhibits a linear dependence on the 2D wave number $q$. We consider the
example of  the Be(0001) surface, with no $d$ bands, which presents a
very-highly populated two-dimensional Shockley surface-state band of the
$s$-$p_z$ symmetry \cite{pl1,pl2,note0}, and as such is an ideal candidate
to investigate the nature of the dynamical screening in a 2D electron gas that
is immersed in a 3D system. The novel result of our detailed band-structure
calculations, in which transitions between the surface-state band and all
other states of the semi-infinite metal are fully included, is that in a real
metal surface where a partially occupied quasi-2D surface-state band {\it
coexists} in the same region of space with the underlying 3D continuum, a
well-defined surface acoustic plasmon exists.

\begin{figure}
\includegraphics[angle=90,width=0.45\textwidth,height=0.3375\textwidth]{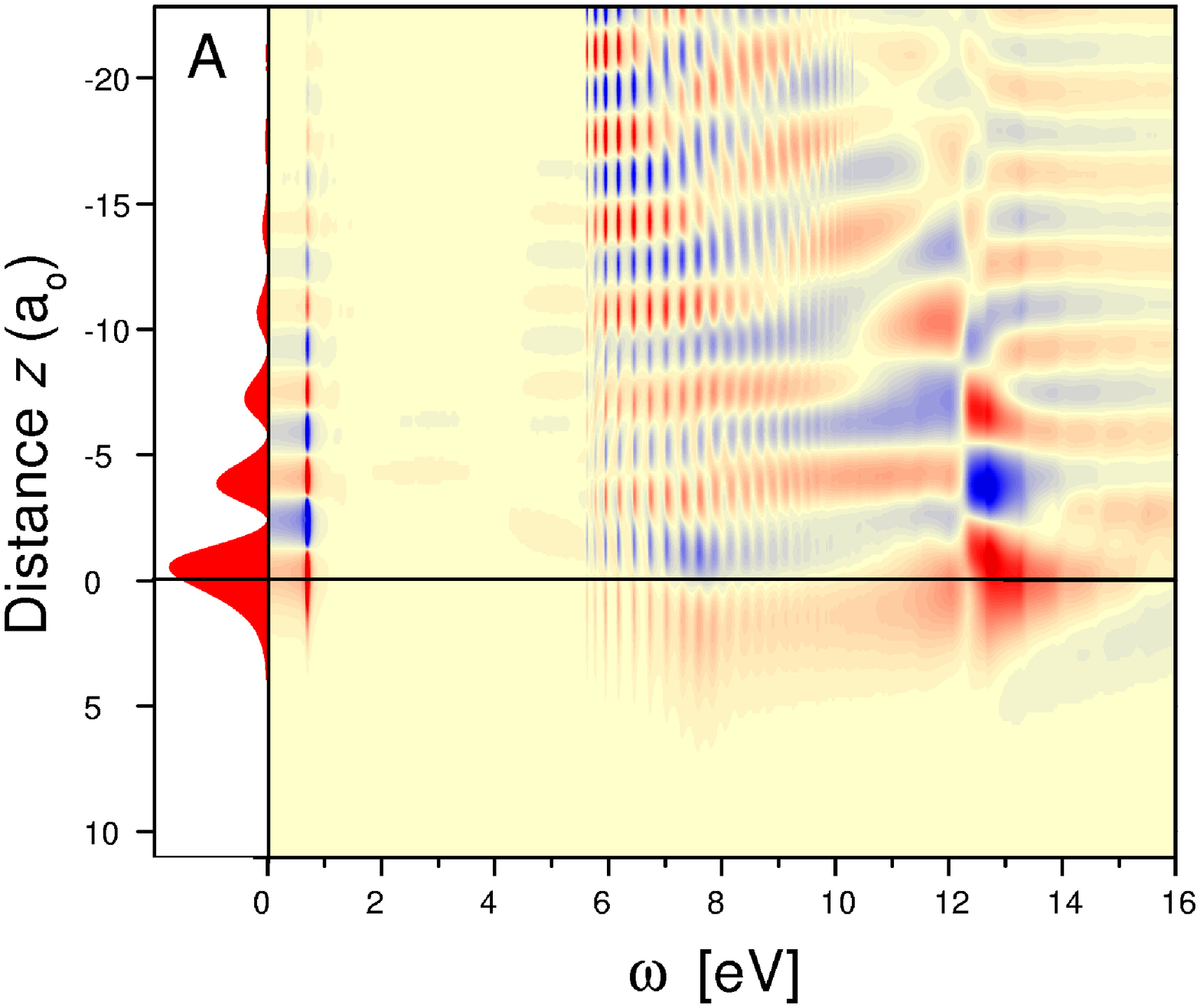}
\includegraphics[width=0.45\textwidth,height=0.3375\textwidth]{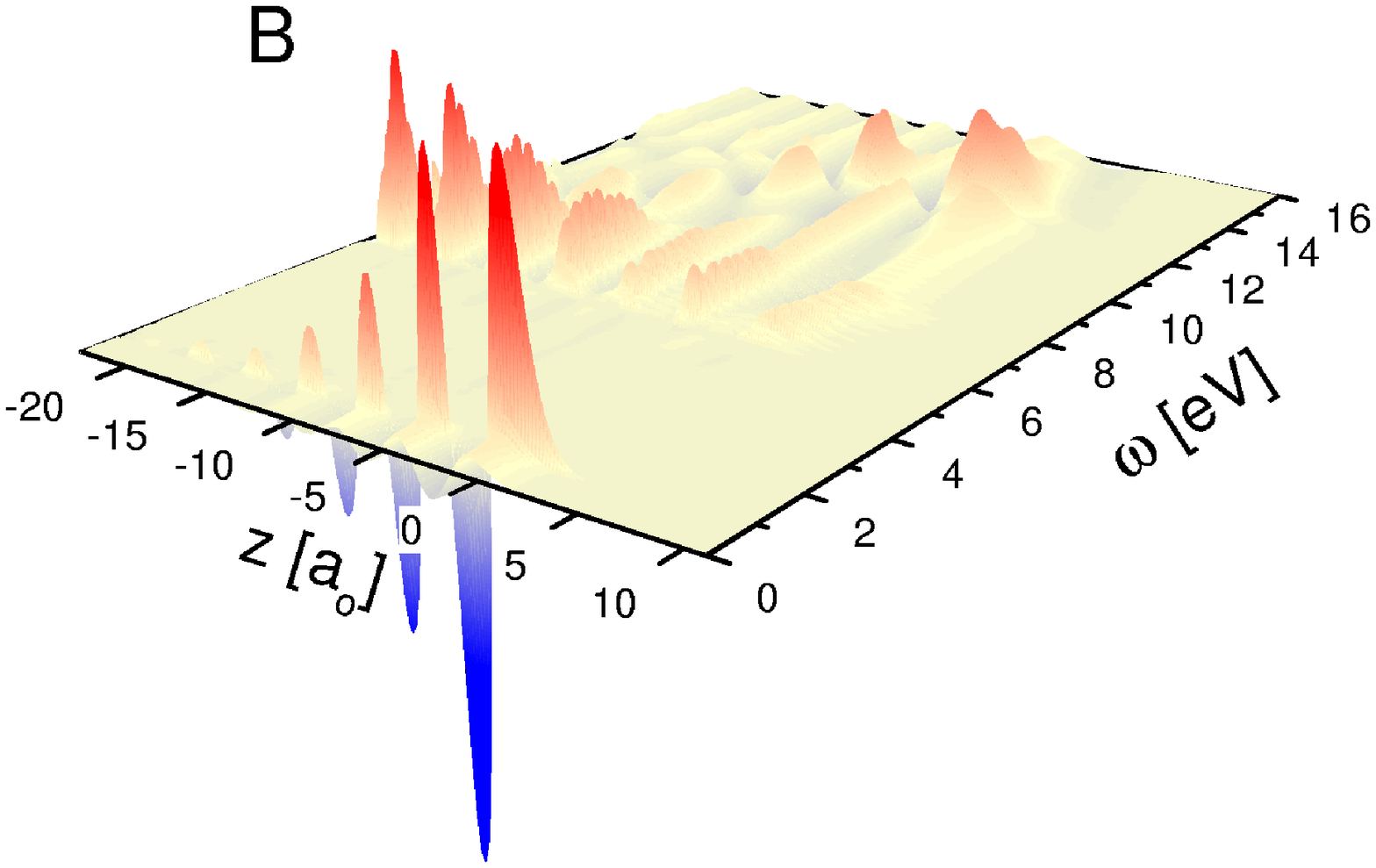}
\caption{RPA calculation of the imaginary part of the electron density
induced in Be(0001), as obtained from eqs. (\ref{deltan}) and (\ref{pot})  for
the wave number $q=0.05$ versus the energy $\omega$
and the coordinate $z$ normal to the surface. The crystal edge ($z=0$) is
chosen to be located half a lattice spacing beyond the last atomic layer, and
$z<0$ corresponds to the interior of the solid. (a) 2D plot of ${\rm Im}[\delta
n(z;q,\omega)]/\omega$. Positive and negative values are represented by red
and blue colours, respectively, more intense colours corresponding to larger
absolute values. The probability density of the Shockley surface state is
represented on the left-hand side (red shaded areas), as a function of the $z$
coordinate. (b) 3D plot of ${\rm Im}[\delta n(z;q,\omega)]/\omega$. }
\label{fig1}
\end{figure} 

First of all, we calculate the eigenfunctions and eigenvalues of a
one-dimensional single-particle hamiltonian that describes the main features
of the surface band structure \cite{review1,review2}. Then we evaluate the
dynamical density-response function $\chi^0(z,z';q,\omega)$ of non-interacting
electrons, and solve an integral equation to obtain the interacting
density-response function $\chi(z,z';q,\omega)$ in the random-phase
approximation (RPA) which is known to be exact in the $q\to 0$ limit
\cite{book}. $\chi(z,z';q,\omega)$ contains both bulk and surface
states and accounts for all possible transitions between them.

Assuming translational invariance in the plane of the
surface, the Fourier components of the induced density satisfy the equation
\begin{equation}\label{deltan}
\delta n(z;q,\omega)=\int
dz'\,\chi(z,z';q,\omega)\,\phi^{ext}(z';q,\omega),
\end{equation} 
and the collective oscillations created by an external potential of the form
\begin{equation}\label{pot}
\phi^{ext}(z;q,\omega)=-(2\pi/q)\,{\rm e}^{qz}
\end{equation}
can be traced to the peaks of the imaginary part of the surface response
function \cite{persson,liebsch}
\begin{equation}\label{g}
{\rm Im}[g(q,\omega)]=\int dz\,{\rm e}^{qz}\,{\rm Im}[\delta n(z;q,\omega)].  
 \end{equation}

In fig.~\ref{fig1} we show our full RPA calculation of the
imaginary part of the electron density $\delta n(z;q,\omega)$ induced in a
Be(0001) surface by the external electrostatic potential of eq. (\ref{pot})
with $q=0.05$. This figure exhibits two distinct
spectral features. The first feature occurs at the surface-plasmon frequency
of valence ($2s^2$) electrons in Be ($\hbar\omega_s\sim 13\,{\rm eV}$), where
${\rm Im}[\delta n]$ exhibits a strong maximum at the surface ($z=0$) and
pronounced Friedel oscillations in the interior of the solid ($z<0$); these
are the main characteristics of the ordinary monopolar surface-plasmon mode,
which would also be present in a semi-infinite free-electron gas (FEG) of bulk
density equal to that of valence electrons in Be. In addition, however, very
localized local maxima are visible near the surface at an energy of $0.6\,{\rm
eV}$, which correspond to a low-energy collective excitation that would be
absent in a semi-infinite FEG. A comparison with the probability density of
the partially occupied Shockley surface state, which is shown by red shaded
areas on the left-hand side of fig.~\ref{fig1}a, clearly indicates that the
low-energy collective excitation originates from this 2D surface-state band.
Nevertheless, such a 2D electron gas {\it alone} would only support a plasmon
that for $q=0.05$ has energy $\omega_{2D}=2.7\,{\rm eV}$, well above
the low-energy excitation that is visible in fig.~\ref{fig1}a, and it is only
the combination of the strongly localized 2D surface-state band with 3D bulk
states which allows the formation of this new mode. 

\begin{figure}
\includegraphics[width=0.45\textwidth,height=0.3375\textwidth]{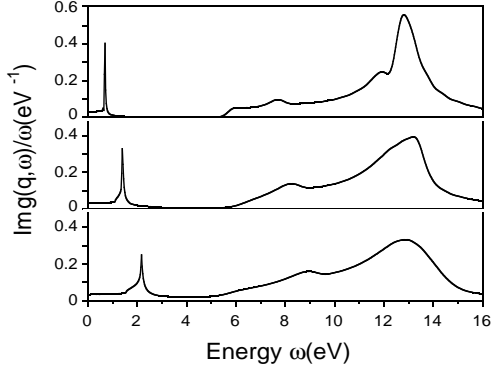}
\caption{Energy-loss function ${\rm Im}[g(q,\omega)]/\omega$ of Be(0001) versus
the excitation energy $\omega$, as obtained from eq. (\ref{g}) for various
values of the wave number: $q=0.05$ (top panel), $0.1$ (middle panel), and
$0.15$ (bottom panel), in units of the inverse Bohr radius $a_0^{-1}$. The
peaks are dictated by the corresponding poles of the surface response function
$g(q,\omega)$. In the long-wavelength limit ($q\to 0$), $g(q,\omega)$ is
simply the total electron density induced by the potential of eq.
(\ref{pot}).} \label{fig2}
\end{figure} 

Figure ~\ref{fig2} shows the imaginary part of the surface-response function
$g(q,\omega)$ of Be(0001), as obtained from eq. (\ref{g}) for increasing values
of $q$. As follows from the figure, the excitation spectra  is dominated by the
conventional surface plasmon at $\hbar\omega_s\sim 13\,{\rm eV}$, which can be
traced to the characteristic pole that the surface response function
$g(q,\omega)$ of a bounded 3D electron gas exhibits at this energy, and a
well-defined low-energy peak with {\it linear} dispersion. Figure ~\ref{fig2}
clearly shows that the low-frequency mode has linear dispersion, with the
sound velocity being very close to the 2D Fermi velocity $v_F^{2D}$.

\begin{figure}
\includegraphics[width=0.45\textwidth,height=0.3375\textwidth]{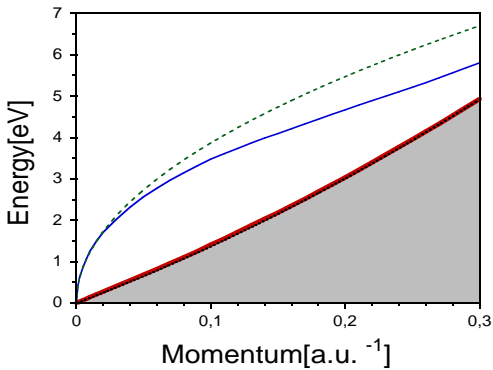}
\caption{Dispersion of the new low-energy collective mode that is
visible in fig.~\ref{fig2} (red solid line), together with the result we
obtain when bulk states are omitted in the evaluation of the energy-loss
function (blue solid line) and the conventional plasmon dispersion
$\omega_p^{2D}=v_F^{2D}\sqrt{q}$ of electrons in a purely 2D
electron gas with $v_F^{2D}=0.45\,v_0$ (green dashed line). The dotted
line represents the upper edge $\omega_u^{2D}=v_F^{2D}q+q^2/2$ of
the 2D e-h pair continuum (shaded area). We note that momentum and
energy conservation prevents 2D e-h pairs to be produced for energies above
$\omega_u^{2D}$.} \label{fig3}
\end{figure}  

In fig.~\ref{fig3} we show the energy of the low-energy mode in Be(0001) versus
$q$ (red solid line), as derived from the maxima of our calculated surface loss
function ${\rm Im}[g(q,\omega)]$, together with the well-defined plasmon
energies that we obtain when only the surface-state band is considered in the
evaluation of the dynamical density-response function (blue solid line). While
the plasmon energies of electrons in the {\it isolated} surface-state band
nicely reproduce in the long-wavelength limit the conventional plasmon
dispersion $\omega_{2D}$ of electrons in a 2D electron gas (green dashed
line), the combination of this surface-state band with the underlying  3D
system yields a {\it new} distinct mode whose energy lies just above the upper
edge $\omega_u^{2D}=v_F^{2D}q+q^2/2$ of the 2D e-h pair
continuum (shaded area). At small $q$, this surface {\it acoustic} mode has
linear dispersion, with the sound velocity being very close to the Fermi
velocity of the 2D surface-state band.

Figures ~\ref{fig1}-\ref{fig3} show that acoustic surface plasmons are
low-energy collective oscillations appropriate to very localized slow
Shockley electrons with Fermi velocity $v_F^{2D}<v_0$ plus their associated
cloud of swift bulk electrons with  $v_F^{3D}\sim v_0$. These excitations
cannot occur at the characteristic energy of the usual 2D plasmons, where
Landau damping would be possible by virtue of 3D e-h pair creation. Instead,
an acoustic plasmon mode occurs at energies that follow closely the upper edge
of the 2D e-h pair continuum.

In conclusion, we have shown that the combination of a localized surface-state
2D band of slow electrons with a bulk 3D electron gas of fast electrons
($v_F^{3D}>v_F^{2D}$), which takes place in the (0001) surface of Be, leads
to the formation of a {\it novel} low-energy collective excitation. This new
{\it acoustic} mode has linear dispersion at small wave vectors, with the
sound velocity being very close to the Fermi velocity of the 2D surface-state
band. A similar mode is also found to exist at the (111) surfaces of the noble
metals \cite{silkin}. We hope that our prediction of acoustic surface
plasmons will stimulate new experiments in this direction.

Partial support by the University of the Basque Country, the Basque
Unibertsitate eta Ikerketa Saila, the Spanish MCyT, and the Max Planck
Research Award Funds is gratefully acknowledged.


\begin{thebibliography}{100}
\bibitem{note0} On Be(0001), a Shockley surface state with a binding energy
of $2.8\,{\rm eV}$ at $\bar\Gamma$ exists in the $\Gamma_3^+-\Gamma_4^-$ bulk
band gap. This surface state disperses with momentum parallel to the surface,
thereby forming a surface-state band with a 2D Fermi energy
$E_F^{2D}=2.8\,{\rm eV}$. If one takes the effective mass of surface-state
electrons to be equal to the free-electron mass $m_e$, the 2D Fermi velocity
is found to be $v_F^{2D}=0.45\,v_0$, with $v_0=e^2/\hbar$ being the Bohr
velocity.
\end{thebibliography}
\end{document}